\documentstyle[10pt]{amsart}

\begin{document}
\title[] {The Friedmannian model \\ of our observed Universe}

\author[]{Vladim\'{i}r Skalsk\'{y}}

\address {Faculty of Materials Science and Technology of the Slovak Technical University,
917 24 Trnava, Slovakia }

\email{skalsky@@mtf.stuba.sk, Fax: ++421-805-27454}

\maketitle

\begin{abstract} According to observations, in our Universe for
gravitational phenomena in a Newtonian approximation the Newtonian
non-modified relations are valid. The Friedmann equations of universe
dynamics describe infinite number of relativistic universe models in
Newtonian approximation, but only in one of them the Newtonian non-modified
relations are valid. From these facts it results that the Universe
is described just by this only Friedmannian universe model with Newtonian
non-modified relations.\end{abstract} \medskip

According to observations -- realized in the course last five centuries --
in our {\it expansive homogeneous and isotropic relativistic Universe\,} for
gravitational phenomena in Newtonian approximation the Newtonian non-modified
relations are valid.

{\it The Friedmann general equations of homogeneous and isotropic relativistic
universe dynamics\,} [1] describe infinite number of homogeneous and isotropic
relativistic universe models in Newtonian approximation, but only in one
of them the Newtonian non-modified relations are valid [2].

From these facts it results unambiguously that our observed Universe is
described just by this only Friedmannian model of universe with Newtonian
non-modified relations.

The deductive-reductive determination of the Friedmannian model of universe
with Newtonian non-modified relations you can see in [3].

The Friedmannian model of universe with Newtonian non-modified relations,
i.e. Friedmannian model of {\it the (flat) expansive non-decelerative (homogeneous
and isotropic relativistic) Universe (with total zero and local non-zero
energy)} (ENU) [4] is a special partial solution of the Friedmann general
equations of homogeneous and isotropic relativistic universe dynamics:
\begin{equation}\label{1a}
\dot{a}^2=\frac{8\pi G\rho a^2}{3}-kc^2 +
\frac{\Lambda a^2 c^2}{3}, \tag{$1a$}
\end{equation}
\begin{equation} \label{1b}
2a\ddot{a}+\dot{a}^2= -\frac{8\pi G p a^2}{c^2}-kc^2+
\Lambda a^2 c^2,\tag{1b}
\end{equation}
with $k = 0$, $\Lambda = 0$ and a state equation [5]:
\setcounter{equation}{1}
\begin{equation}
 p=-\frac13 \varepsilon ,
\end{equation}
where $a$ is the gauge factor, $\rho$ is the mass density, $k$ is the curvature
 index, $\Lambda$ is the cosmological member, $p$ is the pressure, and
 $\varepsilon$ is the energy density.

 The Friedmann equations (1a) and (1b) with $k = 0$, $\Lambda = 0$ and the
 state equation (2) determine the parameters of Friedmannian model of ENU,
 which we express -- for better transparency -- in all possible variants [2],
 [3]:
 \begin{equation}
 a=ct=\frac{c}{H}=\frac{2Gm}{c^2}=\sqrt{\frac{3c^2}{8\pi G \rho}},
 \end{equation}\begin{equation}
 t=\frac{a}{c}=\frac{1}{H}=\frac{2Gm}{c^3}=\sqrt{\frac{3}{8\pi G \rho}},
 \end{equation}\begin{equation}
 H=\frac{c}{a}=\frac{1}{t}=\frac{c^3}{2Gm}=\sqrt{\frac{8\pi G \rho}{3}},
 \end{equation}\begin{equation}
 m=\frac{c^2 a}{2G}=\frac{c^3 t}{2G}=\frac{c^3}{2GH}=\sqrt{\frac{3c^6}
 {32\pi G^3\rho}},
 \end{equation}\begin{equation}
 \rho=\frac{3c^2}{8\pi G a^2}=\frac{3}{8\pi G t^2}=\frac{3H^2}{8\pi G}=
 \frac{3c^6}{32\pi G^3 m^2}=-\frac{3p}{c^2},
 \end{equation} \begin{equation}
  p=-\frac{c^4}{8\pi G a^2}=-\frac{c^2}{8\pi G t^2}=-\frac{c^2 H^2}{8\pi G}=
  -\frac{c^8}{32\pi G^3 m^2}=-\frac{c^2 \rho}{3}=-\frac{1}{3} \varepsilon ,
 \end{equation}
where $t$ is the cosmological time, $H$ is the Hubble ``constant", and $m$
is the mass of ENU.

From relations (3)--(8) it results that the parameters of ENU are mutually
linearly linked. For fundamental parameters of  ENU are valid
relations [2], [6]:
\begin{equation} m=Ca=Dt,\end{equation}
where $C$ and $D$ are (total) constants:
\begin{equation}
C=\frac{m}{a}=\frac{m}{ct}=\frac{Hm}{c}=\sqrt{\frac{8\pi G\rho m^2}{3c^2}}=
\frac{c^2}{2G}=6.734\,67(15)\times 10^{26}\hbox{kg\,m}^{-1},
\end{equation}\begin{equation}
D=\frac{cm}{a}=\frac{m}{t}=Hm=\sqrt{\frac{8\pi G\rho m^2}{3}}=
\frac{c^3}{2G}=2.019\,00(37)\times 10^{35}\hbox{kg\,s}^{-1}.
\end{equation}

The matter-space-time properties of ENU [7] you can see in [6].
\bigskip\medskip

\begin{center}{\bf References}\end{center}\medskip

{
\makebox[5mm][l]{[1]}\parbox[t]{12cm}{Friedmann, A. A.: {\em Z. Phys.} {\bf 10}, 377-386 (1922);
{\bf 21}, 326-332 (1924).}

\makebox[5mm][l]{[2]}\parbox[t]{12cm}{Skalsk\'{y}, V.: {\em DYNAMICS OF
THE UNIVERSE in the Consistent and
Distinguished Relativistic, Classically-Mechanical and Quantum-Mechanical
Analyses}, Slovak Technical University, Bratislava, 1997. \\
Information about
this book can received on e-mail address: {\tt skalsky\symbol{64}mtf.stuba.sk}}

\makebox[5mm][l]{[3]}\parbox[t]{12cm}{Skalsk\'{y}, V.:
{\em Deductive-reductive determination of the Universe model,}
{\tt http://www.mtf.stuba.sk/$\sim$skalsky/}}

\makebox[5mm][l]{[4]}\parbox[t]{12cm}{Skalsk\'{y}, V.: In: J.
Dubni\v{c}ka (ed.): {\em Philosophy, Natural Sciences and
Evolution} (Proceedings of an Interdisciplinary Symposium, Smolenice,
December 12 -- 14, 1988), Slovak Academy of Sciences, Bratislava, 1992,
pp. 83--97.}

\makebox[5mm][l]{[5]}\parbox[t]{12cm}{Skalsk\'{y}, V.: {\em Astrophys. Space Sci.}
{\bf 176}, 313--322 (1991) (Corrigendum:
{\bf 187}, 163 (1992).).}

\makebox[5mm][l]{[6]}\parbox[t]{12cm}{Skalsk\'{y}, V.: {\em The matter-space-time
properties of our observed Universe,}
{\tt http://www.geocities.com/ResearchTriangle/Thinktank/8273/}}

\makebox[5mm][l]{[7]}\parbox[t]{12cm}{Skalsk\'{y}, V.: Astrophys. Space Sci.
{\bf 219}, 275--289 (1994).}
}

\end{document}